\documentstyle[12pt]{article}
\newcommand{\be}{\begin{equation}}
\newcommand{\ee}{\end{equation}}
\def\n{\noindent}
\catcode `\@=11
\catcode `\@=12

\title{\bf\huge A duality relation : global monopole and texture}

\author{Naresh Dadhich\thanks{E-mail : nkd@iucaa.ernet.in} \\
{\sl Inter-University Centre for Astronomy \& Astrophysics,}\\
{\sl Post Bag 4, Ganeshkhind, Pune - 411 007, India.} 
} 
\date{}
\begin{document}
\maketitle

\begin{abstract}

We  resolve  the  entire  gravitational  field;i.e.  the  Riemann 
curvature  into its electric and magnetic parts. In general,  the 
vacuum   Einstein  equation is symmetric in  active  and  passive 
electric parts. However it turns out that the Schwarzschild solution,
which is the unique spherically symmetric vacuum solutions  can  be characterised by a  slightly  more  general 
equation which is not symmetric. Then the duality transformation, 
implying interchange of active and passive parts will relate  the 
Schwarzschlid particle with the one with global monopole  charge. 
That  is the two are dual of each-other. It further turns out  that  flat 
spacetime is dual to massless global monopole and global texture spacetimes.

\end{abstract}

\n PACS numbers : 04.20,04.60,98.80Hw

\newpage 

Like  the  electromagnetic  field,  and  following  the  analogus 
procedure,  gravitational field;i.e. Riemann curvature tensor can 
as  well be resolved into electric and magnetic parts relative  a 
unit  timelike  vector [1-2].Then the Einstein  equation  can  be 
written in terms of electric and magnetic parts of the curvature. 
It  was  noticed  that the Ricci and the  Einstein  tensors  were 
related  through a duality relation in electric  (interchange  of 
active  and  passive)  parts of the  Riemann  curvature  and  its 
implication  on  matter content of spacetime has  been  discussed 
[2].It  turns out that under the duality transformation in  which 
active  and  passive electric parts interchange  their  roles,  a 
fluid  solution  remains a fluid solution with  its  density  and  
pressure   transforming,   while  energy  flux   and   trace-free 
anisotropic pressure remaining unaltered. \\

In this note we wish to investigate spherically symmetric  vacuum 
spacetime  in relation to the duality transformation. In  general 
the  vacuum Einstein equation is symmetric in active and  passive 
electric  parts.  The question is, can we have an  equation  that 
distinguishes  between active and passive parts and yet giving  a 
vacuum solution? If yes, then it would be interesting to  enquire 
what happens then under duality transformation? In particular for 
spherical  symmetry,  the Schwarzschild solution  is  the  unique 
vacuum solution. Can we obtain the Schwarzschild solution from an 
equation that distinguishes between active and passive parts? The 
answer  is yes, we can obtain the solution from a  slightly  more 
general  equation  that is not symmetric in  active  and  passive 
parts.  Now what solution follows when duality transformation  is 
performed?  Interestingly it turns out that the solution dual  to 
the  Schwarzschild  describes  the  Schwarzschild  particle  with 
global  monopole  charge  [3].Similarly  it  can  be  shown  that 
spacetime dual to flat describes a  massless global monopole or a 
uniform relativistic potential [4-5]. That is duality simply  puts 
on a global monopole charge on spacetime.\\

Global  monopoles are supposed to have been created during  phase 
transition   in   the  early  Universe  [6-7]. They   are   stable 
topological   defects   produced   when   global   symmetry    is 
spontaneously broken. In particular for a global monopole, it  is 
$O(3)$  broken into $U(1)$. A solution for a  Schwarzschild  particle 
with  global  monopole charge has been obtained by  Barriola  and 
Vilenkin  [3].By employing the generalised equation we show  that 
the Schwarzschild and the Barriola-Vilenkin solutions are related 
through  the duality transformation;i.e. they are dual  of  each-
other.  Like  the  Schwarzschild solution,  the  global  monopole 
solution  [3]  is  also unique.  Further  the  other  topological 
defect,  global  texture  spacetime [8-9]  is  an  isotropic  and 
homogeneous  solution  of  the equation  obtained  under  duality 
transformation  from the one characterising flat spacetime.  That 
is  global texture is a homogeneous dual, while  massless  global 
monopole  or uniform potential is static dual of flat  spacetime. 
Applications  to  cosmology and properties  of  global  monopoles 
[7,11-13] and of global textures [8-9,14-18] have been studied by 
several authors.\\

Let  us  begin  with  the resolution  of  the  Riemann  curvature 
relative to a unit timelike vector, as follows :

\be
E_{ac} = R_{abcd} u^b u^d,  \tilde E_{ac} = *R*_{abcd} u^b u^d
\ee

\be
H_{ac} = R*_{abcd} u^b u^d = H_{(ac)} - H_{[ac]}
\ee

\n where

\be
H_{(ac)} = C*_{abcd} u^b u^d
\ee

\be
H_{[ac]} = \frac{1}{2} \eta_{abce} R^e_d u^b u^d.
\ee

\n Here $c_{abcd}$  is  the  Weyl conformal  curvature, $\eta_{abcd}$
is  the  4-dimensional volume element. $E_{ab} = E_{ba}, {\tilde E}_{ab}
= {\tilde E}_{ba}, (E_{ab}, {\tilde E}_{ab}, H_{ab})
u^b = 0,~ H= H^a_a = 0$ and $u^a u_a = 1$. 
The Ricci tensor could then be written as

\be
R^b_a = E^b_a + {\tilde E}^b_a + (E + {\tilde E}) u_a u^b -
{\tilde E } g^b_a + \frac{1}{2} \eta_{amn} H^{mn} g^b_0
\ee

\n where $E = E^a_a$        and  $\tilde E = \tilde E^a_a$. 
It may be noted that $E = (\tilde E + \frac{1}{2} T)/2$            defines 
the  gravitational  charge density  while ${\tilde E}= - T_{ab}
u^a u^b$            defines  the 
energy density relative to the unit timelike vector $u^a$  .The vacuum 
equation, $R_{ab} = 0$ is in general equivalent to  

\be
E ~ or ~ {\tilde E} = 0,~ H_{[ab]} = 0 = E_{ab} + {\tilde E}_{ab}
\ee

\n while for spherical symmetry, what would suffice, is

\be
H_{[ab]} = 0 = {\tilde E},~ E_{ab} + {\tilde E}_{ab}
= \lambda g^{1}_a g^{1}_b
\ee

\n where $\lambda$    is a scalar. Here we have taken the resolving vector to 
be  hypersurface orthogonal. It can be easily verified  that  for 
the metric

\be
ds^2 = c^2(r,t) dt^2 - a^2(r,t) dr^2 - r^2 (d \theta^2 + sin^2 \theta
d \varphi^2)
\ee

\n the  equation (7) characterises the Schwarzschild solution.  What 
really  happens is that ${\tilde E} = 0$ gives $a = (1-2M/r)^{-1/2}$                , while  the 
other  equation  yields $ac = 1$ and $\lambda=0$. Note that eqn.(7) 
implies $\lambda = 0$, then it becomes equivalent to the vacuum
equation (6). That is  the  slightly 
more  general equation (7) than the vacuum equation (6)  is  good 
enough  for characterisation of vacuum in this case, because  the 
Schwarzschild solution is the unique vacuum solution in spherical 
symmetry. \\

\n We consider the duality transformation,

\be
E_{ab} \longleftrightarrow {\tilde E}_{ab}, ~H_{ab} \longleftrightarrow
H_{ab}
\ee

\n which  implies $R_{ab} \longleftrightarrow G_{ab}$. 
This is due to the  fact  that $R_{ab}$     
and $G_{ab}$ are  defined by contraction of  Riemann  and  its 
double  dual [19]. Let us now employ this duality  transformation 
to eqn.(7).That means

\be
H_{[ab]} = 0 = E,~ E_{ab} + {\tilde E}_{ab} = \lambda g^{1}_a
g^{1}_b.
\ee

\n Its general solution for the metric (8) is given by 

\be
c = a^{-1} = (1 - 2k - \frac{2M}{r})^{1/2}.
\ee

\n This is the Barriola-Vilenkin solution [3] for the  Schwarzschild 
particle with global monopole charge $\sqrt {2k}$. Here $E=0$
gives $c^{\prime} = const. a/r^2$, and the other equation 
would imply  $ac = 1$, which will  yield $c = (1-2k - 2M/r)^{1/2}$                   and $\lambda = 2k/r^2$. This  has  non-zero 
stresses given by

\be
T^0_0 = T^1_1 = \frac{2k}{r^2}.
\ee

\n A    monopole    is    described    by    a    triplet    scalar,                    $\psi^a (r) = \eta f(r) x^a/r, x^a x^a = r^2$,
which  through  the usual Lagrangian  generates  energy-momentum 
distribution at large distance from the core exactly of the  form 
given  above  in (12) [3].  Like the Schwarzschild solution the monopole solution (11) is also the unique solution of eqn.(10). \\

\n On the other hand, flat spacetime is characterised by

\be
{\tilde E}_{ab} = 0 = H_{[ab]}, E_{ab} = \lambda g^{1}_a g^{1}_b
\ee

\n leading to $c=a=1$. Its dual will be 

\be
E_{ab} = 0 = H_{[ab]}, {\tilde E}_{ab} = \lambda g^{1}_a g^{1}_b
\ee

\n yielding the general solution,

\be
c^{\prime} = a^{\prime} = 0 \Longrightarrow c=1, a = const. = (1-2k)^{-1/2}
\ee

\n which is non-flat and represents a global monopole of zero  mass, 
as it follows from the solution (11) when $M=0$. \\

\n It  can be readily shown that the above solution  corresponds  to 
uniform relativistic potential [4-5].It turns out that the vacuum 
equation  for  the  metric (8) ultimately  reduces  to  the  usual 
Laplace    equation   and   its   first   integral.    That    is                 
$R^0_1 = 0$ and $R^0_0 = R^1_1$ lead to $c^2 = a^{-2} = 1+
2 \phi(r)$, then we finally write[5],

\be
R^0_0 = -\bigtriangledown^2 \phi = 0
\ee

\be
R^2_2 = R^3_3 = - \frac{2}{3} (r \phi)^{\prime} = 0.
\ee

\n The    solution   of   the   first   equation   is    well-known,             
$\phi = -k-M/r$ while  the second equation determines $k=0$, 
which was  free  in 
the Newtonian theory. This is the crucial difference between  the 
Newtonian and the relativistic treatment of this problem. Clearly          
$\phi = -k \not= 0$ is not a vacuum solution and it would generate stresses precisely 
of the form given above in (12).Thus the massless global monopole 
metric (8) with (15) can as well be looked upon as the  spacetime 
describing  a uniform relativistic potential, $\phi = -k \not= 0$.
It may be  noted 
that  the uniform potential spacetime is purely supported by  the 
passive  electric  part.It has been argued and  demonstrated  [5] 
that passive part accounts for non-linear character of the field. 
That is non-linearity produces spatial curvature, which does  not 
vanish  even when potential is set equal to a constant  different 
from  zero. Since this uniform potential unlike the  Newtonian  theory 
has  non-zero  physical  attribute  in  GR  and  hence  it  would 
represent  a  purely  relativistic spacetime  with  no  Newtonian 
analogue. \\

\n Further it  turns  out that a perfect fluid with $\rho + 3p = 0$ 
goes  to 
flat spacetime under the duality transformation [20].This is  the 
equation  of  state,  which means $E=0$,  characterising  global 
texture  [18]. That  is, the necessary condition  for  topological 
defects;  global  textures  and monopoles is   $E = 0$.  For  the 
isotropic  and homogeneous FRW metric, this would  determine  the 
scale factor $S(t) = \alpha t + \beta, $ and $\rho = 3
(\alpha^2 +$ k) $/ (\alpha t + \beta)^2, $ k $= \pm 1, 0$.
It is in fact the homogeneous and isotropic 
solution  of eqn.(14) which is  also  unique. 
Its static analogue was the massless global monopole (15). That is
how global monopole and global texture spacetimes are dual to flat
spacetimes. \\
 
\n Further  it  turns  out  that the  spacetime  with $E=0$  
can  be 
generated   by  considering  a  hypersurface   in   5-dimensional 
Minkowski space defined, for example, by 

\be
t^2 - x^2_1 - x^2_2 - x^2_3 - x^2_4 = k^2 (t^2 - x^2_1 - x^2_2 - x^2_3)
\ee

\n which consequently leads to the metric

\be
ds^2 = k^2 dT^2 - T^2 [d \chi^2 + sinh^2 \chi (d \theta^2 + sin^2 \theta 
d \varphi^2)].
\ee

\n Here $T^2 = t^2 - x^2_1 - x^2_2 - x^2_3 $ and $\rho = 3(1-k^2)/k^2 T^2$.                               The above construction  will  
generate spacetimes of global monopole, cosmic strings (and their 
homogeneous versions as well), and global texture like  depending 
upon  the  dimension  and character of the  hypersurface,  and  of 
course, $E=0$   always;i.e.  zero   gravitational   mass 
[10].The  trace of active part measures the gravitational  charge 
density,   responsible  for  focussing  of  congruence   in   the 
Raychaudhuri  equation  [21].The topological  defects thus have 
vanishing focussing density. \\

\n It is remarkable that the Schwarzschild solution with and without 
global  monopole  charge  could  be  related  through  a  duality 
transformation, which implies interchange of roles of active  and 
passive  electric parts of the filed. In general the empty  space 
equation  is  symmetric  in active and passive  parts  and  hence 
duality  will not make any difference. In spherical symmetry,  we 
have been able to characterise the unique Schwarzschild  solution 
by an equation which is not symmetric, and hence it gives rise to 
new  solution under duality. This is how  Schwarzschild  particles 
with  and without global monopole are shown to be dual  to  each-
other under the duality transformation. Their mass-free  versions 
will  relate  flat spcetime with zero mass  global  monopole  (or 
uniform    potential)   spacetime.   Note   that   the    duality 
transformation amounts to adding or removing global  monopole 
charge from spacetime.\\

\n The  crucial point for the duality transformation to work  is  to 
break  the  symmetry  between active and  passive  parts  in  the 
equation. In whichever situations this could happen, the  duality 
transformation would work to define duality relationship  between 
spacetimes.  Note that $E$ remains unchanged under the duality
transformation which means it does 
not  affect the active gravitational charge, implying  invariance of gravitational  acceleration on free particles. 
It is interesting to note that the  symmetry  in 
eqn.(7)  has been broken by adding something on the right,  which 
is made  up  of  a product of a spacelike  vector  pointing  in  the 
(radial)  direction  of acceleration. This seems  to   suggest  a 
general procedure to break symmetry; i.e. add on the right of (7) 
term  which  is  made  up of spacelike  vector  pointing  in  the 
direction  of the acceleration. Here we have taken  the  timelike 
resolving vector to be hypersurface orthogonal which is
suggested and is consistent with the symmetry and physics
of the situation. This procedure may 
however  not  always  lead to a  viable  application  of  duality 
transformation. The crucial thing to happen is that the  equation  
distinguishes  between  active  and  passive  parts. \\

\n Application of the duality transformation, apart from the  vacuum 
case  considered  here, has been considered for  fluid  spacetime 
[20].  The duality transformation could be easily applied to  the 
eletrovac equation including the $\Lambda$-term. Here the analogue
of the master equation (10) is

\be
H_{[ab]} = 0, E = \Lambda - \frac{Q^2}{2r^4}, ~ E^b_a + 
{\tilde E}^b_a = (- \frac{Q^2}{r^4} + \lambda) g^1_a g^b_1
\ee

\n which has the general solution $c^2 = a^{-2} = (1-2k-2M/r
+ Q^2/2r^2 - \Lambda r^2/3)$ and $\lambda = 2k/r^2$. The
analogue of eqn(7) will have ${\tilde E} = - \Lambda - Q^2/2r^4$
instead of $E$ in (20). Thus the duality transformation works in general
for a charged particle in the deSitter Universe.
The details will be given  elswhere  (2). The next question  we  wish  to 
address in future investigation is to find in which symmetries
does the duality work? 
In other words this will be equivalent to asking which other 
symmetry  admits global monopole charge and  global texture,  and 
in  general  topological  defects? One of  the  most  interesting 
questions will  be to find a global monopole version of  the  Kerr 
solution.  That  is  to  put a  global  monopole charge  on  a  rotating 
particle.  This is a hard problem which we have been  working  on 
for  quite  a while without success. Hopefully the duality might
throw some new light leading to the solution. At any  rate  for  spherical 
symmetry  it  does provide some deeper insight, which  is  in  itself  interesting 
enough. \\

\n Finally massless global monopole or uniform potential, and global 
texture  spacetimes  are  shown to be  dual  to  flat  spacetime,  
should they be thought of as ``minimally curved''? \\

\n {\bf Acknowledgement  :}  I  wish to thank Jose  Senovilla  for  useful 
correspondence,  and  LK Patel and Ramesh  Tikekar  for  fruitful 
discussions. \\

\end{document}